\ifdefined\SingleColumn
\documentclass[12pt, draftclsnofoot, onecolumn, a4paper]{IEEEtran}
\else
\documentclass[article, a4paper]{IEEEtran}
\fi

\usepackage{graphicx}
\usepackage{units}
\usepackage{array}
\usepackage{multirow}
\usepackage{multirow}
\usepackage{rotating}
\usepackage{textcomp}
\usepackage[cmex10]{amsmath}
\usepackage{enumitem}
\usepackage{color}
\usepackage{anyfontsize}
\usepackage{textcomp}
\usepackage{todonotes}
\usepackage{amsfonts}
\usepackage{arydshln}
\usepackage{ulem}
\usepackage{xcolor}

\usepackage{lipsum}
\usepackage[pscoord]{eso-pic}

\newcommand{\placetextbox}[3]{
  \setbox0=\hbox{#3}
  \AddToShipoutPictureFG*{
    \put(\LenToUnit{#1\paperwidth},\LenToUnit{#2\paperheight}){\vtop{{\null}\makebox[0pt][c]{#3}}}%
  }%
}%

\newcommand{\TSwReqAPI}{T_{\mathit{sw\_req}}^{\mathit{API}}}
\newcommand{\TSwReqDRVind}{T_{\mathit{sw\_req}}^{\mathit{DRV\_ind}}}
\newcommand{\TSwReqDRVdep}{T_{\mathit{sw\_req}}^{\mathit{DRV\_dep}}}
\newcommand{\TSwReq}{T_{\mathit{sw\_req}}}
\newcommand{\TSDMACReq}{T_{\operatorname{SDMAC\mathit{\_req}}}}
\newcommand{\TSDMACCon}{T_{\operatorname{SDMAC\mathit{\_con}}}}
\newcommand{\THw}{T_{\mathit{hw}}}
\newcommand{\THwReq}{T_{\mathit{hw\_req}}}
\newcommand{\THwCon}{T_{\mathit{hw\_con}}}
\newcommand{\TSwConISR}{T_{\mathit{sw\_con}}^{\mathit{ISR}}}
\newcommand{\TSwConDRVdep}{T_{\mathit{sw\_con}}^{\mathit{DRV\_dep}}}
\newcommand{\TSwConDRVind}{T_{\mathit{sw\_con}}^{\mathit{DRV\_ind}}}
\newcommand{\TSwConAPI}{T_{\mathit{sw\_con}}^{\mathit{API}}}
\newcommand{\TSwCon}{T_{\mathit{sw\_con}}}
\newcommand{\TPc}{T_{\operatorname{SDMAC}}}
\newcommand{\TTr}{T_{\mathit{toa}}}
\newcommand{\TPath}{T_{\mathit{path}}}

\newcommand{\THwStar}{T^*_{\mathit{hw}}}

\newcommand{\TSwConDRVdepStar}{T_{\mathit{sw\_con}}^{\mathit{* DRV\_dep}}}

\newcommand{\TTrStar}{T^*_{\mathit{toa}}}

\begin{document}
\placetextbox{0.5}{1}{This is the author's version of an article that has been published in this journal.}
\placetextbox{0.5}{0.985}{Changes were made to this version by the publisher prior to publication.}
\placetextbox{0.5}{0.97}{The final version of record is available at \href{https://dx.doi.org/10.1109/TII.2018.2873205}{https://dx.doi.org/10.1109/TII.2018.2873205}}%
\placetextbox{0.5}{0.05}{Copyright (c) 2018 IEEE. Personal use is permitted.}
\placetextbox{0.5}{0.035}{For any other purposes, permission must be obtained from the IEEE by emailing pubs-permissions@ieee.org.}%

\title{SDMAC: A Software-Defined MAC for Wi-Fi to Ease Implementation of Soft Real-time Applications\thanks{This work was partially supported by Regione Piemonte and the Ministry of Education, University, and Research of Italy in the POR FESR 2014/2020 framework, Call ``Piattaforma tecnologica Fabbrica Intelligente'', Project ``Human centered Manufacturing Systems'' (application number 312-36). Copyright (c) 2018 IEEE. Personal use of this material is permitted. However, permission to use this material for any other purposes must be obtained from the IEEE by sending a request to pubs-permissions@ieee.org. The authors are with the National Research Council of Italy, Istituto di Elettronica e di Ingegneria dell'Informazione e delle Telecomunicazioni (CNR-IEIIT), I-10129 Torino, Italy (e-mail: \{name.surname\}@ieiit.cnr.it).}}

\author{Gianluca~Cena, \textit{Senior Member}, \textit{IEEE}, Stefano~Scanzio, \textit{Member}, \textit{IEEE}, and\\Adriano~Valenzano, \textit{Senior Member}, \textit{IEEE}}

\maketitle

\begin{abstract}
In distributed control systems where devices are connected through Wi-Fi,
direct access to low-level MAC operations may help applications to meet their timing constraints.
In particular, the ability to timely control single transmission attempts on air, 
by means of software programs running at the user space level, 
eases the implementation of mechanisms aimed at improving communication timeliness and reliability.
Relevant examples are deterministic traffic scheduling, seamless channel redundancy, rate adaptation algorithms, and so on.

In this paper, a novel architecture is defined, we call SDMAC,
which in its current embodiment relies on conventional Linux PCs equipped with commercial Wi-Fi adapters.
Preliminary SDMAC implementation on a real testbed and its experimental evaluation 
showed that integrating this paradigm in existing protocol stacks constitutes a viable option,
whose performance suits a wide range of applications characterized by soft real-time requirements.
\end{abstract}

\IEEEpeerreviewmaketitle


\section{Introduction}

\mbox{Wi-Fi} \cite{2016-std-80211} adoption in industrial scenarios has been steadily increasing over the past years.
This is mainly due to its high throughput and complete interoperability with Ethernet,
which achieve ubiquitous connectivity between devices, 
reducing at the same time wiring harness complexity.
However, when distributed real-time control systems are taken into account where devices are interconnected 
through \mbox{Wi-Fi}, simple and flexible mechanisms are typically required for configuring and managing network stations.
They permit to easily define, develop, and test new effective applications and solutions.
This is witnessed, for instance, by the recent introduction of software-defined wireless networks (SDWN) \cite{2017-IEEEaccess-SDWN}.

In order to meet the specific timeliness and reliability requirements of factory automation systems,
parameters of the medium access control (MAC) and physical (PHY) layers may have to be suitably tuned, 
even at runtime.
In particular, it should be possible for applications  to manage frame transmission on air with finer detail than typical allowed in \mbox{Wi-Fi} 
(where, e.g., the retransmission process is completely handled in hardware by adapters and hidden to the users).

The ability of the sender to timely start a frame transmission on air 
and to timely obtain a notification of the delivery outcome
enables deterministic overlays to be layered atop \mbox{Wi-Fi},
which help making transmission latencies known in advance and, as much as possible, bounded.
Moreover, applications may be interested in receiving 
management frames, like beacons, which typically are under control of the driver and are not made available to the user, or in accessing information hidden in specific registers of the \mbox{Wi-Fi} adapter.

Currently, many commercial \mbox{Wi-Fi} adapters rely on a SoftMAC architecture, where most functions of the 
MAC sublayer management entity (MLME)  are implemented in software by the device driver 
and are executed by the CPU of the host computer \cite[p.~28]{2010-book-SON}. 
Only time-critical MAC operations (e.g., managing timeouts, like interframe spaces and backoff, and performing the related actions upon their expiry) are executed in hardware by the adapter.
On the contrary, adapters that comply to the FullMAC architecture directly implement the whole IEEE 802.11 protocol stack (both the MAC and the MLME).
Thus, they are more complex and expensive.

Unlike FullMAC, advanced customization is possible for SoftMAC devices, 
which includes redefining protocol parameters and bringing changes to the MLME. 
This is particularly true for the Linux operating system,
where device drivers are often open source and can be easily modified.
However, when changes are required to the MAC or PHY layers, which involve operations executed in hardware, other solutions are needed.
In these cases, functions of the wireless adapter can be possibly implemented using a software-defined radio (SDR) \cite{2008-LANMAN-SDR}, typically by employing an FPGA \cite{2017-CC-SDR}. 
By doing so, practically every aspect of the MAC and PHY layers can be customized to comply with design specifications.
Unfortunately, the required effort is considerably high, and the same applies to cost.

In \cite{2017-WFCS-SDMAC} the software-defined MAC (SDMAC) paradigm was first proposed
to provide applications executing in user space finer control on operations of \mbox{Wi-Fi} adapters.
Its performance, in terms of latency, is not expected to match FPGA-based SDR approaches, 
and also the ability to manage adapter behavior is more limited.
However, SDMAC offers a number of benefits.

To foster its adoption, SDMAC was designed as a flexible framework that relies on commercial \mbox{Wi-Fi} adapters and only requires limited and known modifications to device drivers.
In this way, porting to different equipment or dealing with updated driver releases can be accomplished easily and quickly.
According to the SDMAC paradigm, operations related to custom protocol functions are implemented in user space, 
and rely on a suitable application programming interface (API) that exposes the communication primitives provided by \mbox{Wi-Fi} adapters and the related operating parameters.
To ease implementation and testing, diagnostic information are also provided by SDMAC about its inner state.

\begin{table*}[t]
  \caption{Sample prototypes of the main functions included in the SDMAC API (non-exhaustive --- for informational purposes only).}
  \label{tab:SDMAC_functions}

  \centering
  \ifdefined\SingleColumn
  \includegraphics[width=1\columnwidth]{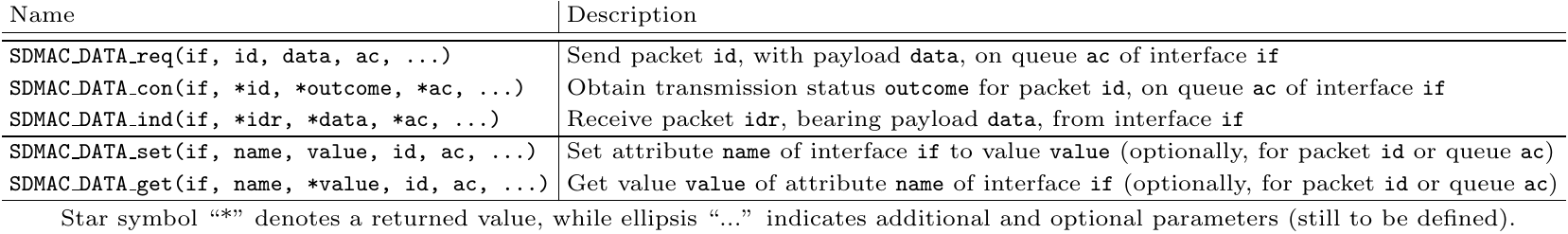}
  \else
  \includegraphics[width=2\columnwidth]{TABLE1_TII-18-0703.pdf}
  \fi
\end{table*}

SDMAC enables direct access to the most basic adapter operations. Probably, the most relevant example are confirmed one-shot transmissions, on which most of the experimental evaluation in this paper focuses.
They can be seen as the elementary building block on which every specific protocol overlay (a supplemental mechanism layered above adapters, and meant to improve some aspect, e.g., determinism) can be built.
In this case, automatic MAC retransmission upon errors is disabled, and packet transmission simply consists of one DATA frame followed, in case of success, by the related ACK frame in the opposite direction.
Having the ACK frame for single attempts managed in the hardware of the adapter 
unburdens the software of this task and ensures higher reliability and performance, 
without practically limiting SDMAC flexibility.
Multi-shot transmissions, that permit a configurable number of retries, can be also defined.

In addition, SDMAC provides a timely notification when transmission ends, which makes the outcome of delivery available to applications in user space.
In this way, the retransmission process can be completely managed in software,
which enables additional behaviors besides those foreseen by the standard specification.
In general, the exact pattern of frames appearing on air, including the related timings,
can be decided in SDMAC according to user-defined rules.

Preliminarily to the stable definition of the SDMAC interface, one must determine if the performance this approach delivers on real devices meets the typical requirements of industrial applications.
In this work, a thorough statistical characterization of the latencies introduced by SDMAC was performed,
in order to assess whether this approach is adequate for the intended scenarios 
or more complex and expensive solutions, like SDR, have instead to be pursued.

Compared to the preliminary proposal in \cite{2017-WFCS-SDMAC}, a number of enhancements are provided in this paper: 
1) a taxonomy of the most important SDMAC service primitives has been sketched;
2) an enhanced measurement system has been developed, where timestamps are acquired in several different places of the protocol stack;
3) a set of specific guidelines and optimizations is provided that sensibly improves SDMAC performance over \cite{2017-WFCS-SDMAC}; and,
4) a more extensive analysis was performed on the dependence between latency and transmission outcome.
The paper is organized as follows:
in Section \ref{sec:SDMAC} the SDMAC architecture, its basic properties, and possible application scenarios are described.
Section \ref{sec:implementation} focuses on SDMAC implementation, while Section \ref{sec:measurement} describes the measurement system.
A discussion on results is included in Section \ref{sec:results}, followed by concluding remarks.

\section{Software-Defined MAC}\label{sec:SDMAC}
SDMAC implementation resides partly in user space and partly in kernel space.
In particular, small blocks of code have to be placed in specific positions of the device driver 
to perform specific operations (e.g., to capture relevant information).
Effortless integration in existing device drivers was a key design requirement.
Part of SDMAC duties is to transfer information from user space to kernel space, and vice-versa.

\subsection{Taxonomy of SDMAC services}
SDMAC services can be subdivided in two broad classes: 
\emph{transmission-oriented} and \emph{MAC-layer management}.

\subsubsection{Transmission-oriented services}
\label{sec:TransmissionOrientedFunctions}

These services deal with data exchanges among STAs at the \emph{data-link} layer, 
and are the building blocks that permit applications to precisely coordinate their actions on air,
enabling timely data delivery.
While they somehow resemble conventional MAC transmission services performed in hardware,
the overall protocol execution takes now place in software under user control.

SDMAC services comply to the OSI model: \emph{request} and \emph{confirm} primitives are defined in the originator, and \emph{indication} in the recipient.
The \emph{response} primitive is not foreseen, because in \mbox{Wi-Fi} it is mapped on ACK frames,
which are automatically sent by the adapter of the recipient STA after a short interframe space (SIFS) has elapsed from DATA frame reception.
For unconfirmed traffic, only the \emph{request} and \emph{indication} primitives are needed.
SDMAC data exchange services are implemented through the \texttt{SDMAC\_DATA\_req()}, \texttt{SDMAC\_DATA\_con()}, and \texttt{SDMAC\_DATA\_ind()} functions.
A conceptual draft of their prototypes, together with the most important parameters, is reported in Table~\ref{tab:SDMAC_functions}. 

\texttt{SDMAC\_DATA\_req()} is used to send a packet. 
Parameter \texttt{if} identifies the target interface in multi-adapter configurations, whereas \texttt{id} is a packet identifier. 
The value of \texttt{id} is initialized by the calling application, and must be unique in the related scope.
If more than one application is invoking SDMAC primitives at the same time, each of them has its own scope and is free to select its specific values for \texttt{id}. 
Coherence between the \texttt{id} values in user and kernel spaces is managed directly by SDMAC. 
In the quite common implementation we considered in this paper, a character device is used to link these two execution contexts.
Parameter \texttt{ac} specifies the queue of the adapter where the packet will be buffered for the forthcoming transmission.
Typically, \texttt{ac} corresponds to one of the $4$ access categories (AC) foreseen by every recent adapter complying with IEEE 802.11e, i.e., \emph{voice}, \emph{video}, \emph{best effort}, and \emph{background}, and permits to differentiate the quality of service (QoS) for specific classes of transmitted packets.

\texttt{SDMAC\_DATA\_con()} is used to obtain a timely notification at the end of packet transmission.
It also provides the transmission outcome (either success, when the ACK frame is received, or failure, in case \emph{ACKTimeout} expired and the retry limit was reached).
To permit confirmations to be paired by the user with the related requests, the \texttt{id} of the packet to which the notification refers is returned. 
Its default behavior is to block the caller until the outcome is provided.

\texttt{SDMAC\_DATA\_ind()} is called by the recipient STA to wait for the arrival of a packet. 
The returned packet identifier \texttt{idr} is unique on the recipient.
Depending on the specific implementation, it could be coupled with the index \texttt{id} chosen by the originating STA (this can be useful, e.g., for the implementation of seamless redundancy \cite{2016-tii-wired}).

\subsubsection{MAC-layer management services}
\label{sec:ManagementFunctions}
These services are meant to complement standard MLME ones, and support the additional features enabled by SDMAC.
MAC implementation is partitioned in several functional blocks, either hardware (in the adapter) or software (in the device driver).
Real adapters (e.g., Atheros) are typically made up of a number of queue control units (QCU) for managing transmission queues, each of which is linked to exactly one DCF control unit (DCU) for dealing with channel access.
Each QCU/DCU pair deals with a specific class of transmissions (e.g., a given AC).
A single protocol control unit (PCU) and a single DMA receive unit (DRU) are also there.

\texttt{SDMAC\_DATA\_get()} and \texttt{SDMAC\_DATA\_set()} are designed bearing in mind the above architecture.
Parameters to be read/written may refer to: 
1) the adapter as a whole (e.g., BSSID, ACK timeout, and general-purpose statistics);
2) a specific queue of the adapter, as specified by \texttt{ac} (e.g., TXOP, AIFSN, CW$^{\operatorname{min}}$ and CW$^{\operatorname{max}}$);
or 3) a single buffered packet, as specified by \texttt{id} (e.g., timestamps).
In the third case, parameters can be also written/read contextually to the related \texttt{SDMAC\_DATA} primitives, by augmenting the related functions with suitable arguments.
Notable examples are the number of allowed tries and their rates (\emph{request}) and the number of actually performed transmission attempts (\emph{confirm}).
To unburden programmers and increase SDMAC performance, default values can be defined for parameters.

A precise definition of the SDMAC API is out of the scope of the current paper, which only focuses on performance aspects, and is left as future work.

\subsection{SDMAC properties}
Several properties make SDMAC appealing for a number of application contexts, with and without real-time constraints.
First, only features required in a particular application context have to be implemented and integrated in the driver, and the same holds for attributes describing the inner MAC status.
For instance, transmission-oriented \texttt{SDMAC\_DATA} functions are only needed when the default send/receive functions provided by the protocol stack do not offer the capabilities required by applications.
In particular, \texttt{SDMAC\_DATA\_con()} is mandatory when notifications on packet delivery are needed.

SDMAC was designed bearing in mind easy porting on a wide range of devices, including updated releases of existing drivers. 
As a matter of fact, a number of projects exist for hard real-time implementations of protocol stacks, but they are seldom up-to-date and hardly work with the most recent network adapters.
Remarkable examples are RTNet and EtherLab.
RTNet provides a hard real-time protocol stack and driver for wired and wireless adapters. 
Unfortunately, \mbox{Wi-Fi} drivers are only available for few legacy adapters and are mostly in a prototype stage.
Likewise, a specific hard real-time driver \cite{2011-TII-EtherCAT} was released for EtherLAB, but it only targets a very specific kind of Ethernet adapter (\mbox{Wi-Fi} is not even envisaged in this case).
While adopting hard real-time drivers is definitely the best choice from a performance viewpoint, it is only acceptable for applications based on specific hardware, for which no updates are foreseen over time.

\begin{figure}[b]
\centering
\includegraphics[width=1\columnwidth]{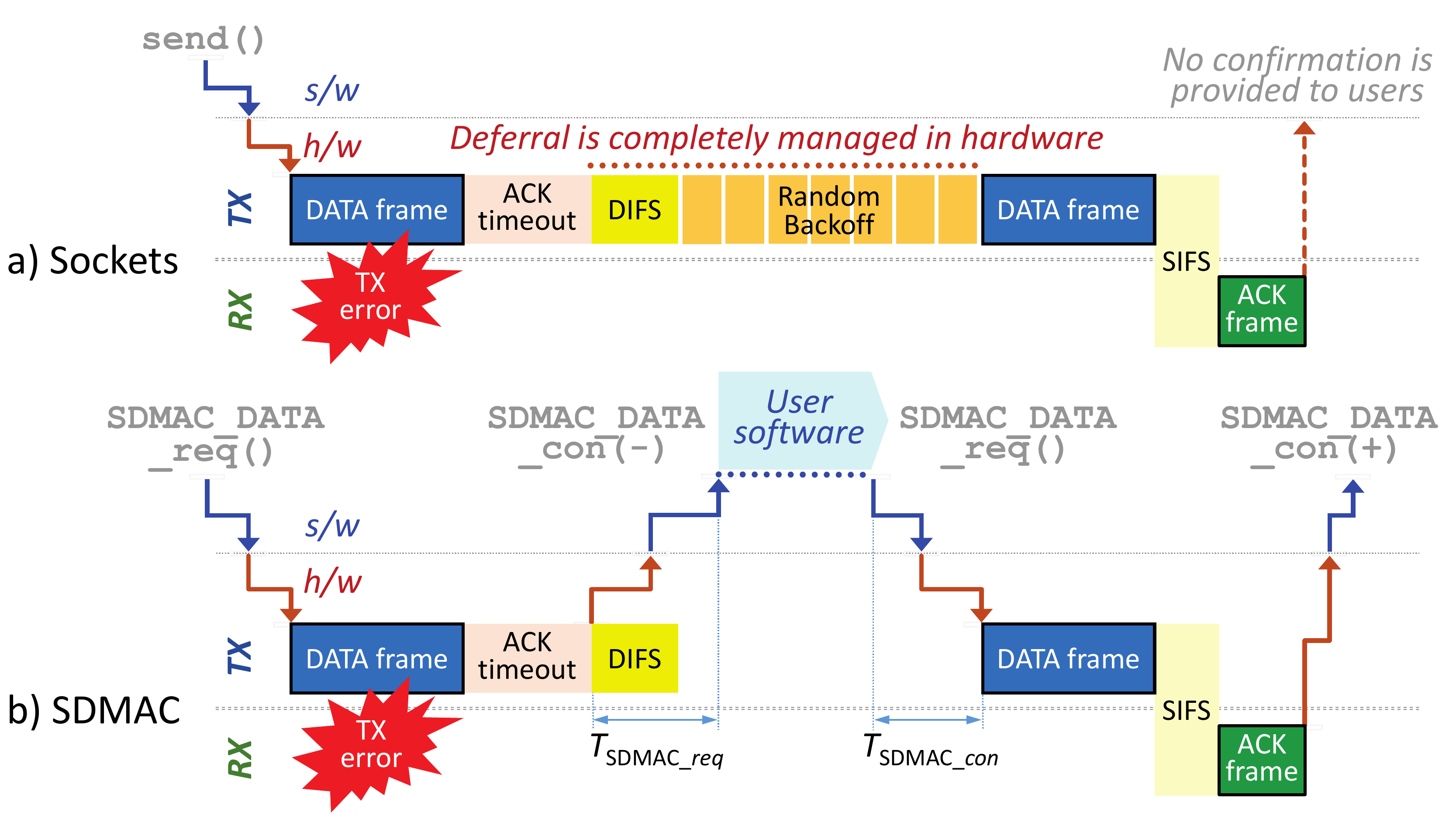}
\caption{Management of retransmissions (conventional sockets and SDMAC).}
\label{fig:sdmac}
\end{figure}

The ability to precisely control IEEE 802.11 MAC behavior by means of code in user space is the most important feature of SDMAC.
It also enables the main \mbox{Wi-Fi} communication parameters to be tuned, even at runtime and with per-packet granularity.
Besides, the availability of a well-defined SDMAC framework may dramatically reduce the time for prototyping wireless systems with specific communication needs.
Developers can benefit from this flexibility, e.g., by defining custom traffic management and enhanced retransmission schemes. 
For example, Fig.~\ref{fig:sdmac} shows how a data transmission is performed, both in conventional \mbox{Wi-Fi} and with a custom SDMAC-based implementation.
When using existing transmission services in Linux (based on sockets), management of retries and backoff is entirely performed by the adapter, and no confirmation is provided to the user when transmission ends.
Conversely, by using SDMAC, it is up to the user to decide the exact way retransmissions are carried out (e.g., how many attempts can be performed and how much they are spaced).

As shown in the figure, code in user space is executed upon failure of transmission attempts, which can drive protocol operations at runtime, enforcing a specific behavior.
The only drawback is that, unlike operations carried out by the adapter with precise timings (tolerances are below $\unit[1]{\mu s}$), delays and jitter introduced by real SDMAC implementations may lower determinism tangibly, which directly affects the benefits it can actually offer. 
In order to prove SDMAC practical feasibility, this paper focuses on the experimental characterization of time-critical functions, and in particular on the use of \texttt{SDMAC\_DATA\_req()} and \texttt{SDMAC\_DATA\_con()}
to perform confirmed one-shot transmissions.
Mainly for space reasons, evaluation of \texttt{SDMAC\_DATA\_ind()} was not performed. 
In the experiments, we measured how much it takes for SDMAC to send a frame on air and obtain the outcome of delivery (kind of a round-trip delay).
This latency is what actually matters for many deterministic MAC overlays,
as sharing the wireless spectrum among STAs can be carried out more efficiently and predictably 
if delays and jitters due to frame originators (not recipients) are known in advance.

It is worth pointing out that using SDMAC to manage retransmissions in software,
as shown in the lower part of Fig.~\ref{fig:sdmac}, negatively affects throughput.
However, this is not the most important performance metric for real-time applications.
In these cases, the ability to precisely control how and when any single attempt, related to a specific frame transmission, takes place on air, is more important.

We decided not to analyze MAC-layer management services because they are not time-critical.

\subsection{Application contexts that can benefit from SDMAC}
\label{sub:application_contexts}

\subsubsection{Time Division Multiple Access}
\label{sec:TimeDivisionMultipleAccess}
Recent \mbox{Wi-Fi} specifications \cite{2016-std-80211} define the enhanced distributed channel access (EDCA) and hybrid-coordination-function controlled channel access (HCCA).
EDCA is completely \emph{distributed}, but relies on a \emph{random} access scheme.
Conversely, while offering \emph{deterministic} access, HCCA relies on a \emph{centralized} coordinator.
The optimal choice in many industrial scenarios, however, is \emph{deterministic distributed} channel access.

Time division multiple access (TDMA) \cite{2013-RTSS-TDMA, 2008-ETFA-TDMA_self, 2011-ICCCN-Vesco, 2014-TMC-Sevani} is a popular distributed approach for sharing the communication support in time-critical applications, and can be applied to wireless networks by constraining each node to access the channel exclusively during its assigned time slots.
If nodes are synchronized, so that they share the same time base \cite{2017-TII-SAUTER-SYNC, 2015-TII-RBIS}, it is possible to coordinate their access to the underlying network even in the absence of a repeated superframe (e.g., in \mbox{Wi-Fi}).
The ability of STAs to send frames on air at precise instants permits to correctly dimension safety margins of time slots.
The more accurate timings are, the lower the wasted bandwidth.
In turn, this means higher process data rate.
Therefore, knowing a priori SDMAC transmission latencies is essential to correctly configure TDMA exchanges.

\subsubsection{Deadline-driven traffic scheduling}
\label{sec:DeadlineDrivenTraffic}
Deciding transmission order of data according to their deadlines is another context where SDMAC can be advantageous.
Besides bounded transmission latency, a prompt notification of the transmission outcome (success or failure) permits the scheduler to timely select the next frame to be sent on air according to specific strategies.
In addition, statistical information collected in the driver and made available to the scheduler via the SDMAC interface enables it to proactively react to environmental changes.
Knowledge of the round-trip delay measured at the user layer for single confirmed transmission attempts can be profitably exploited in scheduling algorithms, e.g., the earliest deadline first (EDF), to account for the actual channel occupation of data exchanges.

For example, in \cite{2017-tii-sched} transmissions (and retries) of pending frames take place according to their absolute deadlines.
The related EDF scheduler can be implemented as a soft real-time application in user space that relies on SDMAC.
Clearly, resulting performance depends on SDMAC overhead and accuracy.
Another relevant example is the SchedWiFi proposal \cite{2015-ETFA-SchedWiFi}, which combines distributed channel access and deadline-driven traffic scheduling and also supports aperiodic traffic.
It is based on TDMA and makes use of a Time-Aware Shaper module to isolate high-priority traffic in predefined slots.

\subsubsection{Seamless redundancy}
\label{sec:SeamlessRedundancy}
SDMAC can be profitably employed to efficiently implement seamless \mbox{Wi-Fi} redundancy \cite{2017-tii-red}. 
The ability to perform confirmed one-shot transmissions, remove packets from transmission queues, and, possibly, abort ongoing transmissions on adapters, enables sophisticated strategies, like duplication avoidance in Wi-Red \cite{2016-tii-wired}, also including proactive heuristics based on network statistics.
A preliminary, simplified implementation is described in \cite{2018-ETFA-RDA}.

\subsubsection{Additional application contexts}
\label{sec:OtherApplications}
Scenarios that can benefit from SDMAC include, e.g., enhanced rate adaptation techniques and real-time roaming of mobile STAs.
Besides, experimental results reported here can be used to characterize in a realistic way in-node delays in network simulators.

\section{SDMAC Implementation}\label{sec:implementation}
\begin{figure}[!t]
\centering
\includegraphics[width=1\columnwidth]{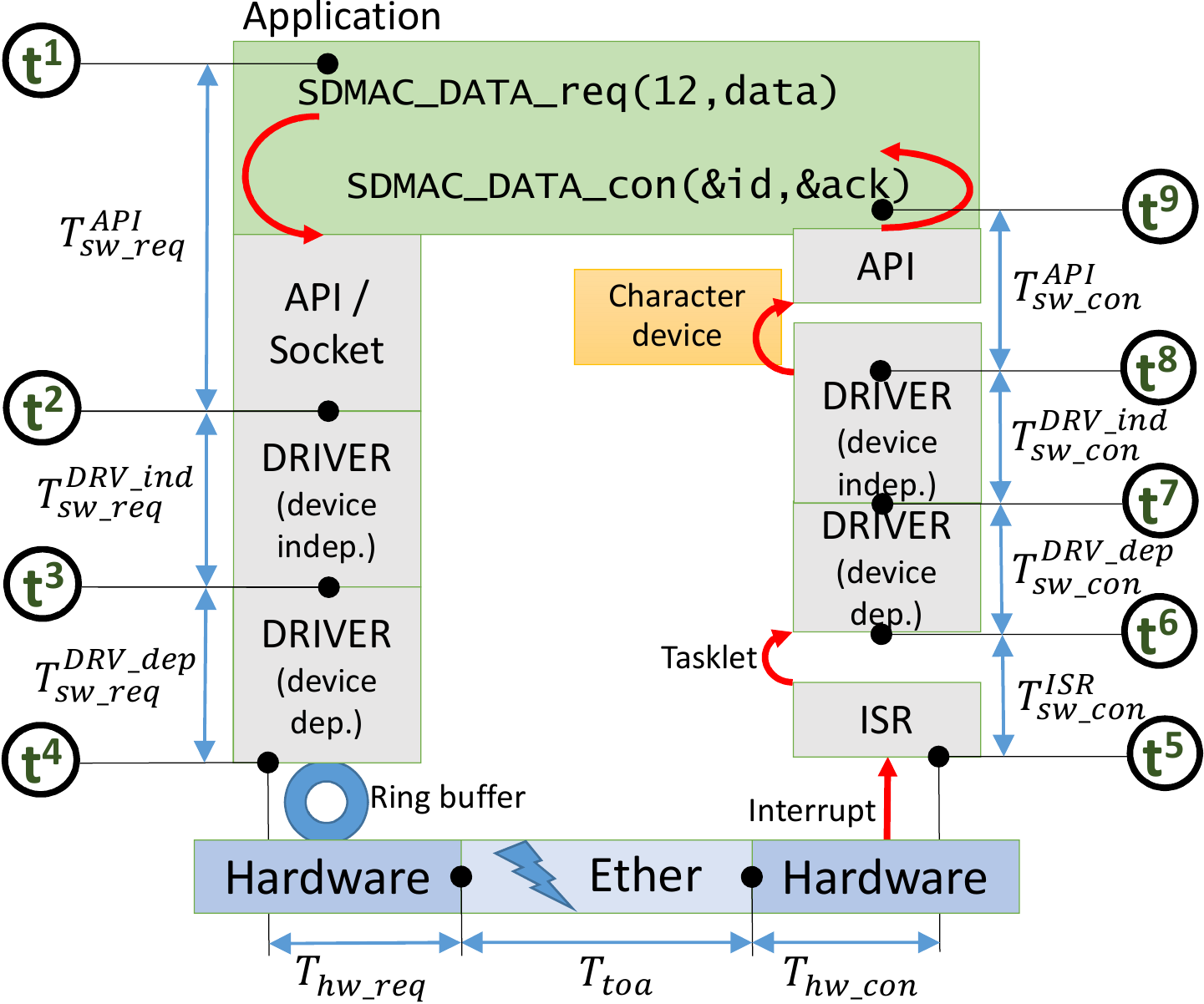}
\caption{SDMAC implementation schema.}
\label{fig:implementation}
\end{figure}

Practical implementation of \texttt{SDMAC\_DATA\_req()} and \texttt{SDMAC\_DATA\_con()} in a typical Linux system (and the related device drivers) is schematically shown in Fig.~\ref{fig:implementation}.

\subsubsection{\texttt{SDMAC\_DATA\_req()}}
A reasonable option is to map the SDMAC primitive for sending data directly on POSIX \emph{raw sockets}, and in particular on the \texttt{sendto()} function, as done in this work, or \texttt{sendmsg()}, if ancillary information have to be conveyed with the packet (e.g., parameters \texttt{id} and \texttt{ac}, as well as attributes managed by \texttt{SDMAC\_DATA\_set()}).
The use of standard POSIX implies that the \emph{request} path (before transmission or air) coincides with the conventional Linux protocol stack.
Consequently, no changes at all have to be made to the driver. 
As shown on the left side of Fig.~\ref{fig:implementation}, the application in user space issues a call to \texttt{SDMAC\_DATA\_req()}, which in turn invokes  \texttt{sendto()}  on the related socket. 
After execution of the \emph{device dependent} and \emph{device independent} components of the driver, the packet is inserted into the \emph{ring buffer}, which coincides with the transmission queue of one of the QCU/DCU blocks of the \mbox{Wi-Fi} adapter.

Implementation of confirmed one-shot transmissions requires automatic MAC layer frame retransmissions to be disabled.
Depending on the specific driver and network adapter, this can be easily achieved in user space
by configuring the retry limit to $0$ (e.g., using the \texttt{iwconfig} command) or, 
if this option is not supported, by modifying the driver (as we did here).
In the latter case, developers are required to find the correct position in the driver code where retransmissions can be disabled.
In the following experiments, Atheros \mbox{Wi-Fi} adapters based on the \texttt{ath9k} driver were used.
They are very popular in the research community, since the driver is available as open source and is not based on proprietary firmware.

For \texttt{ath9k}-compliant adapters, every outgoing packet is associated with a suitable memory structure (\emph{TX descriptor}) that contains packet-specific attributes.
Upon packet transmission request, the related TX descriptor is instantiated and inserted into the ring buffer by the driver. 
It will be fetched autonomously (in DMA) by the network adapter for transmission on air. 
To provide finer control on single transmission attempts, up to four \emph{transmission series} can be defined for each descriptor.
Specific fields exist to configure, e.g., the maximum number of attempts that can be automatically performed for each series (\texttt{tx\_tries0/1/2/3}) and the transmission rate the adapter has to use for them (\texttt{tx\_rate0/1/2/3}).
Actual values for such fields are selected by the driver on a per-packet basis, typically according to the Minstrel \cite{2013-ICC-minstrel, 2017-TII-RATE_VITTURI} algorithm.
Rate adaptation is an effective way to increase communication reliability. 
For custom SDMAC transmission services, e.g., confirmed one-shot transmissions, it has to be implemented at user space level, either by SDMAC or by the application.

\subsubsection{\texttt{SDMAC\_DATA\_con()}}
The originating STA is notified of the outcome of each packet transmission either directly, when an ACK frame is received from the recipient, or indirectly, when it is not received before \emph{ACKTimeout} expiration.
Both events cause the network adapter to raise an interrupt, which is served as soon as possible by the operating system through the related interrupt service routine (ISR) \cite{2014-tii-sauter-isr-delay}. 
The ISR is aimed at managing time-critical tasks.
Instead, other activities are executed as a \emph{tasklet}, which is a Linux mechanism aimed at deferring code execution so as to decrease interrupt response latencies. 
Tasklets are managed by the same CPU that served the interrupt (i.e., that ran the ISR), and are executed in interrupt context (i.e., they cannot be preempted by other tasks).
Tasklet code can be divided in two components, \emph{device dependent} and \emph{device independent}. 
The first makes access to registers of the specific network adapter, while in the latter the same code is shared among all the devices managed by the device driver. 
To make integration of SDMAC into device drivers easier, we placed the code to detect transmission outcomes in the device independent part, to the detriment of latency, which worsens slightly. 
Specifically, for \texttt{ath9k}, it fitted in the \texttt{ieee80211\_tx\_status()} function.

With POSIX sockets, the \emph{confirmation} path does not reach applications directly.
Conversely, the outcome of each packet transmission in SDMAC is transferred in user space by means of a \emph{character device}, and made available to applications through \texttt{SDMAC\_DATA\_con()}.
To this purpose, we used a semaphore in kernel space and a blocking \texttt{read()} system call in the user space \texttt{SDMAC\_DATA\_con()} function.

Determinism of communication between kernel and user spaces could be improved \cite{2012-ETFA-kernel_user}, but this requires to set up a hard real-time environment, which sensibly increases configuration effort. 
Besides, doing so does not practically guarantee hard real-time behavior because, at present, only few prototype implementations exist of hard real-time \mbox{Wi-Fi} device drivers. 
Experimental results highlighted that the jitter induced by the character device is negligible when compared to other components.
A possible alternative to character devices is the \texttt{ioctl()} function, but it requires many more changes to the driver code and was replaced in newer drivers by the \emph{netlink} interface. Unfortunately, \emph{netlink} worsens determinism \cite{2011-ETFA-ioctl}.

\section{Measurement system}\label{sec:measurement}
\begin{table}[t]
  \caption{Description of measurement planes for timestamps.}
  \label{tab:timestamps}
  
  \centering
  \includegraphics[width=1\columnwidth]{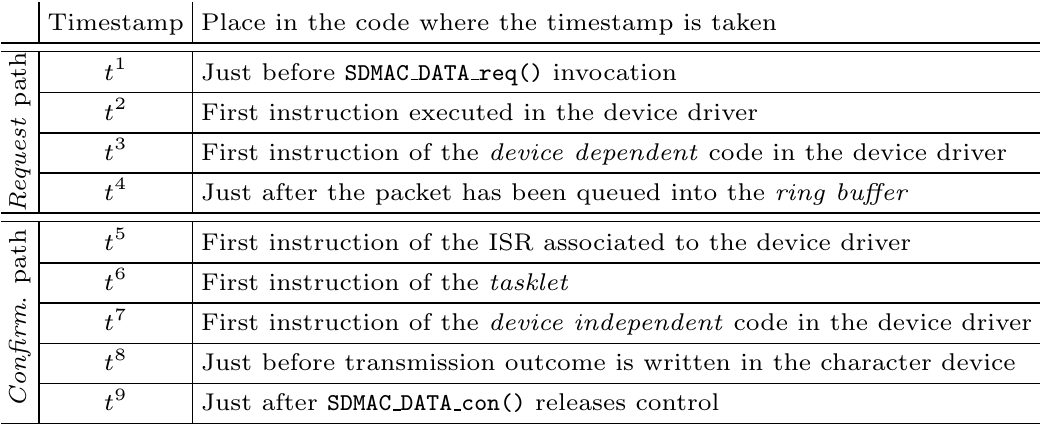}

\end{table}
To profile the most significant software blocks in the \emph{request} and \emph{confirmation} paths,
small, specific pieces of code were added to SDMAC.
As shown in Fig.~\ref{fig:implementation} and Table~\ref{tab:timestamps}, several meaningful reference planes were identified. 
For every packet transmission, a timestamp was obtained as close as possible to each reference plane, by reading the Time Stamp Counter (TSC) register of the CPU. 
The path between the invocation of \texttt{SDMAC\_DATA\_req()} (at time $t^1$) and the unblocking of \texttt{SDMAC\_DATA\_con()} (at time $t^9$) can be decomposed as
\begin{equation}
  \TPath = t^9 - t^1 = \TSDMACReq + \TTr + \TSDMACCon
\end{equation}
where $\TSDMACReq=\TSwReq+\THwReq$ and $\TSDMACCon=\THwCon+\TSwCon$ refer to the delays introduced by SDMAC (both software and hardware) on the request and confirmation paths, respectively, while $\TTr$ is the time taken for transmission on air.
Software latencies on the two paths can be further subdivided as, respectively,
\begin{align}
  \TSwReq & =\TSwReqAPI+\TSwReqDRVind+\TSwReqDRVdep \\
  \TSwCon  & =\TSwConISR+\TSwConDRVdep+\TSwConDRVind+\TSwConAPI . \nonumber
\end{align}

By defining $\THw=\THwReq+\THwCon$ as the overall round-trip delay due to \mbox{Wi-Fi} adapter hardware, the total overhead introduced by SDMAC on confirmed transmission (including both $\TSDMACReq$ and $\TSDMACCon$) comprises all contributions to $\TPath$ in Fig.~\ref{fig:implementation} except $\TTr$ and can be expressed as
\begin{equation}
	\label{eq:TPc}
  \TPc = \TSwReq + \THw + \TSwCon = \TPath - \TTr .
\end{equation}

When automatic MAC retransmissions are disabled, as for confirmed one-shot transmissions, $\TTr$ for packets for which an ACK was successfully received (\emph{acked}) is given by
\begin{equation}
\label{eq:TTr}
\TTr = T_{acc}+\left(T_{\operatorname{DATA}}+T_{\operatorname{SIFS}}+T_{\operatorname{ACK}}\right)
\end{equation}
where $T_{\operatorname{DATA}}$ and $T_{\operatorname{ACK}}$ are the durations of DATA and ACK frames, respectively, $T_{\operatorname{SIFS}}$ is the SIFS duration, and $T_{acc}$ is the time taken by the MAC to access the wireless medium.
$T_{\operatorname{SIFS}}$ is constant and only depends on the specific PHY layer.
When rate adaption algorithms are disabled, also $T_{\operatorname{DATA}}$ and $T_{\operatorname{ACK}}$ are fixed and can be easily computed. 
Instead, $T_{acc}$ is, by its nature, not deterministic, as it depends on the interference due to transmissions performed by nearby wireless devices.

Not every transmitted packet receives a confirmation, as either the DATA or the ACK frame may be corrupted. 
From the originator viewpoint, these situations coincide and denote a transmission failure.
However, in the second case the recipient receives data correctly.
Statistics on latencies for packets for which no ACK was received (\emph{non-acked}) were computed separately. 
The related quantities are identified with superscript ``$^*$'', and not necessarily correspond to acked frames.
An exception is the request path, which is unaffected by the transmission outcome.
For non-acked frames \eqref{eq:TPc} becomes
\begin{equation}
\TPc^*= \TSwReq + \THw^* + \TSwCon^* = \TPath^* - \TTr^*
\end{equation}
where, for confirmed one-shot transmissions,
\begin{equation}
\label{eq:TTrs}
\TTr^* = T_{acc}+T_{\operatorname{DATA}}+T_{\operatorname{ACKTimeout}}.
\end{equation}

\subsection{Software architecture and testbed}
\label{sub:software_architecture}
\begin{figure}[!t]
\centering
\includegraphics[width=1\columnwidth]{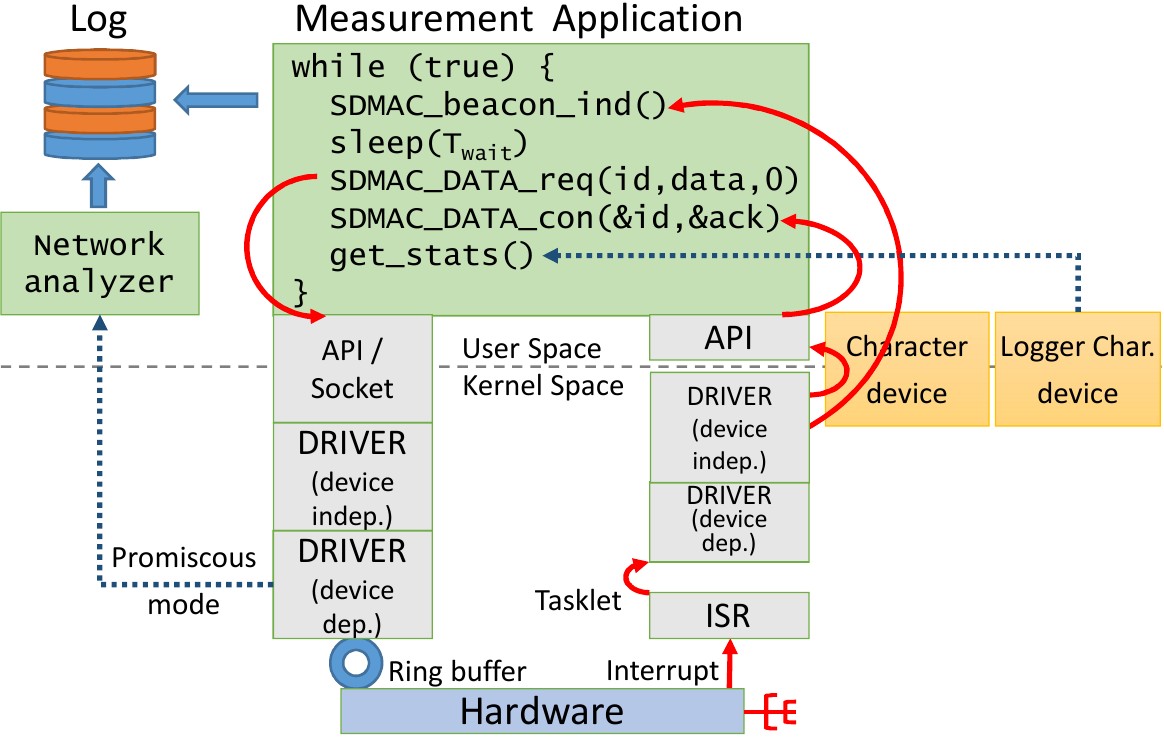}
\caption{Implementation schema of the measurement system.}
\label{fig:measurement}
\end{figure}

Experimental evaluation of SDMAC performance was carried out using a purposely developed testbed, whose architecture is depicted in Fig.~\ref{fig:measurement}.
The measurement system was implemented on a PC equipped with a $\unit[3.5]{GHz}$ Intel\textregistered\ i3-4150 CPU, Intel\textregistered\ B86 Chipset, and $\unit[4]{GB}$ $\unit[1600]{MHz}$ DDR3 Dual Channel RAM, running the Linux kernel v.~3.14.61 and the Ubuntu 14.04.4 LTS distribution.
Energy management features and frequency scaling were disabled to improve determinism,  according to Intel\textregistered\ guidelines \cite{2012-intel-intel-idle, 2017-WFCS-SDMAC}. 
A dual-band TP-Link TL-WDN4800 was used as \mbox{Wi-Fi} adapter, managed by the \texttt{ath9k} device driver v.~4.1.1, and configured to comply with IEEE 802.11a, since this is the simplest way to prevent frame aggregation (the option to explicitly disable aggregation is planned for the final SDMAC version).
A generic access point (AP) was configured to set up an infrastructure \mbox{Wi-Fi} network.
In the context of this paper, its performance is irrelevant, because it only has to reply with an ACK frame to every DATA frame sent by the PC.
Further details on the measurement system can be found in \cite{2017-WFCS-SDMAC}.

\subsection{Measurement technique}
To correctly characterize SDMAC latencies due to the hardware, we need to make $T_{acc}=0$ in the experiments. 
In particular, delays caused by the clear channel assessment, which defers frame transmission when the channel is sensed busy, must be avoided.
To this extent, a peculiar technique was employed, which guarantees that the channel is idle (practically) every time a transmission request is performed by the measurement task.
After $T_{acc}$ contribution has been removed, $\THw$ can be easily evaluated as $\THw = t^5-t^4- \left( T_{\operatorname{DATA}}+T_{\operatorname{SIFS}}+T_{\operatorname{ACK}} \right)$ for acked frames and 
$\THw^* = t^5-t^4- \left( T_{\operatorname{DATA}}+T_{\operatorname{ACKTimeout}} \right)$ for non-acked ones.

\begin{table}[t]
  \caption{Configuration of parameters for experiments}
  \label{tab:parameters}

  \centering
  \includegraphics[width=1\columnwidth]{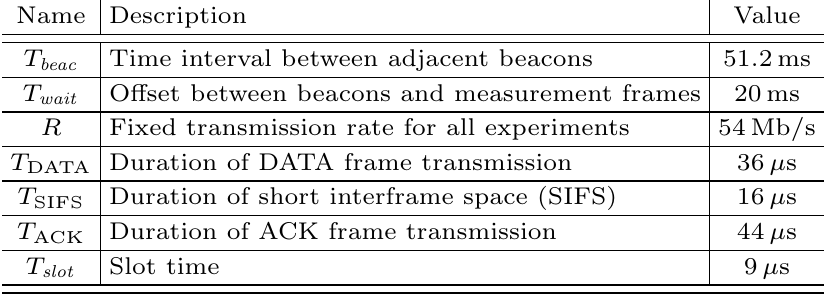}

\end{table}

A number of countermeasures were taken to prevent interference between the frames sent by the measurement task and concurrent traffic on air, as described below.

\subsubsection{Channel selection}
The AP in the testbed (and, as a consequence, the \mbox{Wi-Fi} adapter in the PC) were configured on a channel in the $\unit[5]{GHz}$ band not currently in use by others STAs.
At the time the experiments were carried out, traffic on such band was, on average, by far lower than on the quite crammed $\unit[2.4]{GHz}$ band.
The \texttt{iwlist} Linux command was used to discover a \mbox{Wi-Fi} channel on which no APs were visible in the place where the testbed was deployed.
We found that channel $165$ satisfied this property.
Traffic on channel $165$ was then monitored using WireShark, before and after each experiment. 
To prevent any possible interference with other frames from affecting results, network traffic was logged at runtime by the measurement system (see Section~\ref{sec:network_analyzer}).

\subsubsection{Preventing interference with beacons}
The above countermeasure alone does not solve the problem of interference.
In fact, the AP used in our testbed
periodically broadcasts beacon frames (every $T_{beac}$), which may interfere with the frames sent by the measurement task.
An effective approach to prevent this from happening is to synchronize operations of the measurement task to the AP, so as to evenly interleave measurement frames and beacons.
In particular, the measurement task was instructed to wait for $T_{wait}=\unit[20]{ms} \simeq T_{beac}/2$ following every beacon event before invoking \texttt{SDMAC\_DATA\_req()}.
Function \texttt{SDMAC\_beacon\_ind()}, which relies on the character device, was purposely defined to notify the user application of the beacon arrival (detected in the driver). 
After obtaining the transmission outcome through \texttt{SDMAC\_DATA\_con()}, the measurement task invokes \texttt{get\_stats()}, which relies on a separate character device (\emph{logger}), to transfer timestamps $t^{2...8}$ (acquired in kernel space) to user space, where they are logged to memory.

\subsubsection{Preventing interference with other frames}\label{sec:network_analyzer}
In spite of the previous two countermeasures, the presence of sporadic transmissions on the selected channel can not be ruled out for sure.
Think of, e.g., probe frames sent by mobile phones of people walking around near our testbed. 
This means that some samples may be affected by unexpected access delays.
For them, $T_{acc}$ is not null and unknown.
In order to prevent these events from affecting statistics of SDMAC latencies, such samples have to be identified and discarded.
To this purpose, a specific concurrent thread (\emph{network analyzer}) was run on the PC, which took a timestamp on every frame received on air, with the exclusion of beacons and measurement frames generated by the testbed.
This was accomplished by configuring the network interface in promiscuous mode and using the \texttt{libpcap} library to detect any such frames.

Timestamps of unexpected interfering frames, the $j$-th of which is denoted with $t^{int}_j$, were logged and used in the post analysis phase to filter out samples possibly affected by access delays.
In particular, measured samples where timestamps $t^4$ or $t^5$ fell in any interval $t^{int}_j \pm \unit[10]{ms}$ were discarded.
It is worth pointing out that only $\unit[0.1]{\%}$ of the samples were actually cast away, which confirms the effectiveness of our technique.

\section{Results}
\label{sec:results}
Relevant settings for the experimental campaign are reported in Table~\ref{tab:parameters}. 
In particular, the bit rate was set to $\unit[54]{Mb/s}$ (fixed) to disable the rate adaptation algorithm, hence making the quantities $T_{\operatorname{DATA}}$ and $T_{\operatorname{ACK}}$ constant.
Instead, the beacon interval was shortened from the default value ($\unit[102.4]{ms}$) to $\unit[51.2]{ms}$, to double the number of acquired samples. 
Unless otherwise specified, the payload size of measurement frames was set to $\unit[50]{B}$, which is realistic for process data in industrial scenarios.
Statistical indices on latency samples, which include mean value ($\overline T$), standard deviation ($s_T$), minimum ($T_{\operatorname{min}}$), maximum ($T_{\operatorname{max}}$), and the $99$-, $99.9$- and $99.99$-percentiles ($T_{p99}$, $T_{p99.9}$ and $T_{p99.99}$, respectively), were computed offline from the timestamps acquired in the SDMAC testbed.

\subsection{Interfering load}
The first experimental campaign analyzes to which extent interfering tasks inside the PC affect SDMAC latencies. 
Among the kinds of load considered in \cite{2017-WFCS-SDMAC}, we selected the most aggressive one, that is \emph{I/O load}: 
the \texttt{dd} Linux utility was invoked to transfer, through the hard disk controller, huge amounts of data on the system bus ($\unit[\sim\!\!80]{MB/s}$), which sensibly increases interrupt generation rate ($\sim\!\!215$ interrupts per second). 
On the contrary, the \emph{no load} condition refers to an idle system, where only the tasks of a typical Linux distribution (including the graphical user interface and the \texttt{ssh} server daemon) and the measurement application are running.
Each experiment lasted one day (i.e., $1\,728\,000$ samples), and two configurations were analyzed, \emph{baseline} and \emph{optimized}.

\subsubsection{Baseline}

\begin{table}[t]
  \caption{Experimental results for successful frame transmissions (standard Linux kernel without and with optimizations).}
  \label{tab:std}

  \centering
  \includegraphics[width=1\columnwidth]{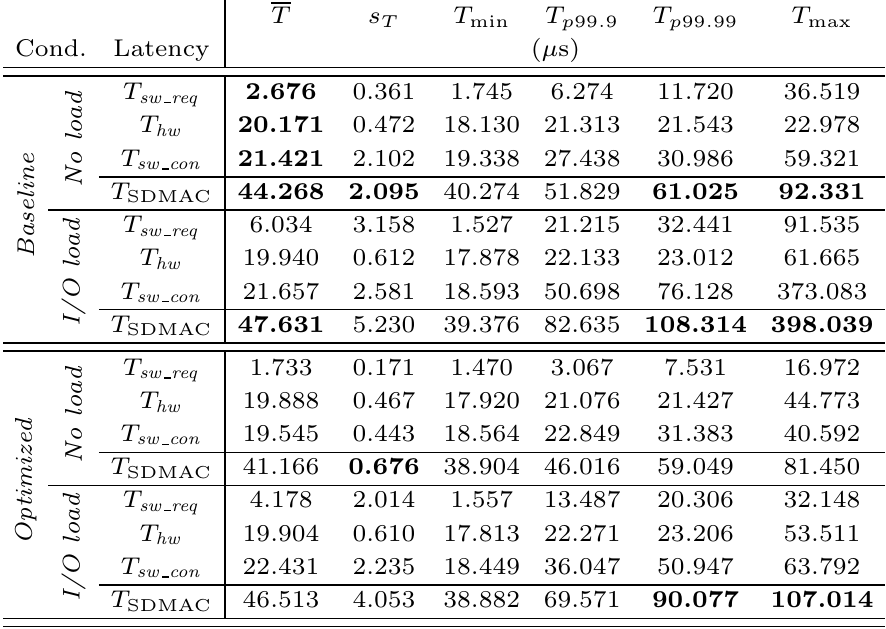}

\end{table}

Standard settings of the Linux distribution were left unchanged, energy management and frequency scaling were disabled \cite{2017-WFCS-SDMAC}, and the Linux kernel was optimized and recompiled for Intel\textregistered\ i3 CPUs (\emph{Core 2/newer Xeon} option in \emph{Processor family} kernel parameter). 
Experimental results for this configuration are reported in the upper part of Table \ref{tab:std}.

Regarding SDMAC overhead ($\TPc$), mean latency in \emph{no load} conditions is $\unit[44.268]{\mu s}$, while real-time statistical indices, $T_{p99.99}$ and $T_{\operatorname{max}}$, are $\unit[61.025]{\mu s}$ and $\unit[92.331]{\mu s}$, respectively. 
All values are bounded and reasonably low. 
In presence of \emph{I/O load}, the average latency is similar, that is $\unit[47.631]{\mu s}$.
Unfortunately, $T_{p99.99}$ and $T_{\operatorname{max}}$ grow to $\unit[108.314]{\mu s}$ and $\unit[398.039]{\mu s}$, respectively.
In this case, the $99.99$-percentile is more than twice the average value, and the maximum, as expected, grew out of control. 
This is no surprise, because SDMAC implementation is not hard real-time.

Considering the overall path of confirmed one-shot transmissions, the biggest contributions to the latency are due to the software confirmation path ($\overline T_{sw\_con}=\unit[21.421]{\mu s}$) and the hardware ($\overline T_{hw}=\unit[20.171]{\mu s}$), which are much higher than the software request path ($\overline T_{sw\_req}=\unit[2.676]{\mu s}$). 
This is good news because, from the application viewpoint, the ability of a node to timely inject a frame on air is more important than the notification latency of transmission outcomes. 
Think, e.g., to scheduled transmissions, including TDMA schemes.

\subsubsection{Optimized}

In this case, some strategies have been devised and applied to increase determinism
and reduce transmission latencies. 
We only considered \emph{general} optimization strategies, which are not targeted to specific devices, brands, or operating system releases. 
All the proposed settings can be applied to conventional PCs and \mbox{Wi-Fi} adapters, on any multi-core CPU running a mainstream Linux distribution. 
Although we re-compiled the Linux kernel, performance only slightly worsens if this part of optimization is skipped. 
Among a relatively high number of strategies we tested, the following ones have been selected.

\begin{figure}[!t]
\centering
\includegraphics[width=1\columnwidth]{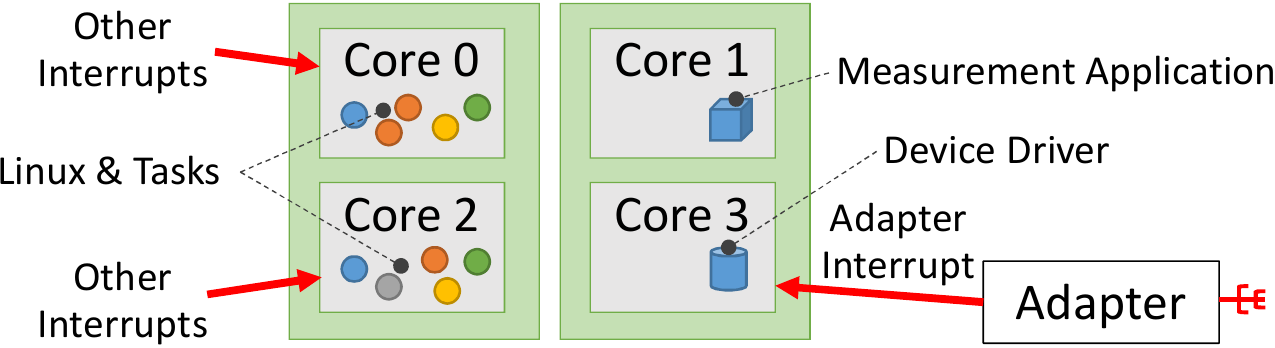}
\caption{Isolation between the execution of SDMAC and other tasks.}
\label{fig:isolation}
\end{figure}

As for the \emph{baseline} configuration, the kernel was compiled with the specific CPU optimizations, but disabling all kernel debugging features.
Regarding optimizations that do not require kernel re-compilation, we partitioned the execution environment (i.e., CPU cores) so as to dedicate one core for the execution of the ``industrial'' application (i.e., the measurement application in the context of this paper), another for the driver, and the remaining ones for the operating system and other tasks (see Fig.~\ref{fig:isolation}). 
The CPU used in our testbed has two physical cores and four logical cores. 
As logical cores are not physically separate entities, results are sub-optimal. 
Our proposed allocation schema can be enforced by setting the default affinity of the operating system and all other tasks to logical cores $0$ and $2$, which reside in the same physical core (by running the kernel with boot parameter \texttt{isolcpus=1,3}). 
Then, the affinity of the kernel thread that executes the device driver was set to core $3$ (with the \texttt{taskset} command), and the priority of the driver was set to the highest-but-one real-time priority with first in, first out scheduling policy (with the \texttt{chrt -f -p 98 <pid>} command). 
To better isolate tasks, we also modified interrupt affinity: all interrupts were scheduled on cores $0$ and $2$, with the only exception of those related to the \mbox{Wi-Fi} adapter, which were scheduled on core $3$ (the same execution core as the device driver). 
Finally, the measurement application was scheduled on the reserved core $1$ by means of the \texttt{taskset} command.

Results for the \emph{optimized} configuration are reported in the lower part of Table~\ref{tab:std}. 
The most significant improvements brought by optimizations regard real-time $\TPc$ indices.
In particular, for the aggressive \emph{I/O load} condition, $T_{p99.99}=\unit[90.077]{\mu s}$ and $T_{\operatorname{max}}=\unit[107.014]{\mu s}$.
All other indices improve as well, including minimum and average values.
Effectiveness of optimizations is also confirmed in \emph{no load} conditions, where determinism is remarkably better. 
For example, standard deviation $s_T$ decreased from $\unit[2.095]{\mu s}$ to $\unit[0.676]{\mu s}$.

\begin{table*}[t]
  \caption{Experimental results for successful/failed transmissions (standard Linux kernel with optimizations --- one-week run).}
  \label{tab:outcomes}

  \centering
  \ifdefined\SingleColumn
  \includegraphics[width=1\columnwidth]{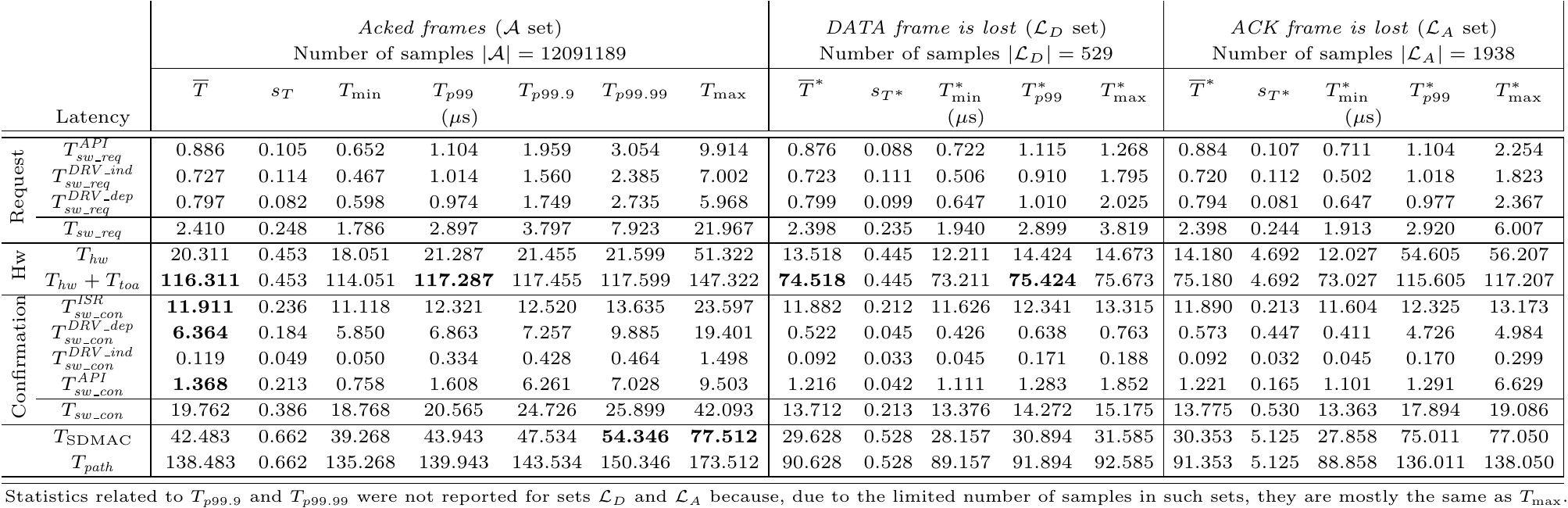}
  \else
  \includegraphics[width=2\columnwidth]{TABLE5_TII-18-0703.pdf}
  \fi
\end{table*}

\subsection{Latency vs. packet size}
\label{sec:packetsize}
\begin{figure}[b]
  \scriptsize
  \centering
  \includegraphics[width=1\columnwidth]{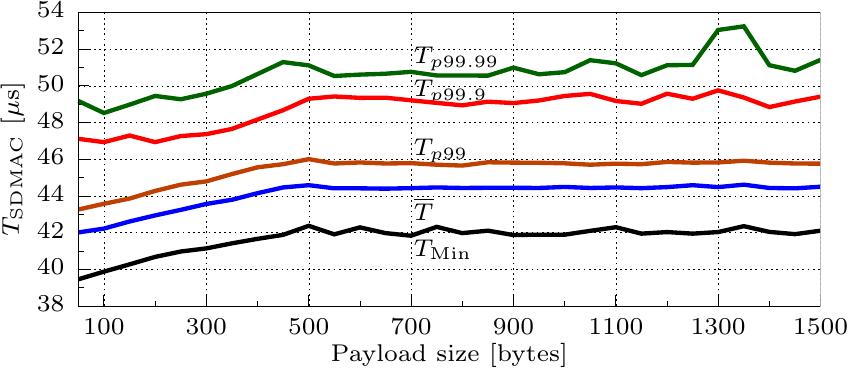}
  \caption{Statistical indices of SDMAC latency vs. payload size.}
  \label{plot:size}
\end{figure}

This experiment was aimed at evaluating the effect of the payload size on latency $\TPc$. 
In practice, the experiment in \emph{no load} conditions with the \emph{optimized} configuration was repeated for different payload sizes (from $50$ to $\unit[1500]{B}$ in $\unit[50]{B}$ steps). 
Each experiment lasted $1$ hour.
In Fig.~\ref{plot:size}, statistical indices of $\TPc$ are plotted.
The shape of curves is piecewise linear. 
For frames up to $\unit[\sim\!500]{B}$, dependence is linear and $\TPc$ increased by $\sim\!\unit[1]{\mu s}$ every $\unit[150]{B}$. 
When payload size is larger than $\unit[500]{B}$, $\TPc$ was stable and practically stayed between $T_{\operatorname{min}}\simeq \unit[42]{\mu s}$ and $T_{p99.9}\simeq\unit[50]{\mu s}$ (as can be seen, the limited duration of experiments made, by necessity, $T_{p99.99}$ not as reliable as other statistics).

Above behavior is due to the fact that, in order to start transmission,
the adapter does not wait for the whole packet to be fetched in DMA from the PC main memory.
Instead, it just acquires a prefix of the packet in advance and loads the remaining part while transmitting on air.

\subsection{Latency vs. transmission outcome}
\label{sec:outcomes}
The last experiment, carried out in typical, realistic operating conditions (\emph{no load}, \emph{optimized} configuration, $\unit[50]{B}$ payload), had two main purposes. 
Firstly, we wanted to analyze SDMAC over a longer period of time, to assess its real-time performance with higher confidence. 
For this reason, the experiment lasted $7$ days.
Secondly, we wished to determine dependence of timings on transmission outcomes.
To this purpose, we split acquired samples into three sets:
\begin{itemize}
\item $\mathcal{A}$:   DATA frame correctly delivered and acked;
\item $\mathcal{L}_D$: DATA frame lost and consequently no ACK frame;
\item $\mathcal{L}_A$: DATA frame correctly delivered but ACK frame lost.
\end{itemize}
A sample is assigned to set $\mathcal{A}$ when \texttt{SDMAC\_DATA\_con()} notifies a successful transmission. 
Instead, when it returns an \emph{ACKTimeout} event, the frame is added to either set $\mathcal{L}_A$, if correctly received by the recipient, or set $\mathcal{L}_D$ otherwise. 
To assign non-acked frames to sets $\mathcal{L}_A$ and $\mathcal{L}_D$, sequence numbers were included in the payload of measurement frames, and a purposely-developed program was run on the recipient side (a virtual AP running on a PC) to log sequence numbers of received frames.
Statistics were evaluated on both the overall latency and individual contributions to the latency.

\subsubsection{Acked frames}
Results for acked frames in set $\mathcal{A}$  are reported in the leftmost part of Table~\ref{tab:outcomes}. 
They confirm the good real-time properties of SDMAC, even over wider time spans.
As can be seen, only $1$ frame out of $10000$ suffered from an internal delay (due to SDMAC overhead) longer than $\unit[54.346]{\mu s}$ (see $T_{p99.99}$ referred to $\TPc$), and the maximum is bounded to a quite low value ($T_{\operatorname{max}}=\unit[77.512]{\mu s}$), despite the system is not hard real-time.

When individual contributions to the SDMAC latency are concerned, most of the time is spent in the $\THw$ and $\TSwCon$ components. 
In particular, the largest part of $\TSwCon$ was caused by the ISR ($\TSwConISR$) and the device dependent part of the driver ($\TSwConDRVdep$), whose mean values were $\unit[11.911]{\mu s}$ and $\unit[6.364]{\mu s}$, respectively.
This means that the time taken by SDMAC to transfer the transmission outcome from the device driver in kernel space to the application in user space ($\TSwConAPI$) is negligible ($\unit[1.368]{\mu s}$, on average). 
Therefore, optimizations aimed at reducing such time further are practically worthless.

\begin{figure*}[t]
  \scriptsize
  \centering
  \ifdefined\SingleColumn
  \includegraphics[width=1\columnwidth]{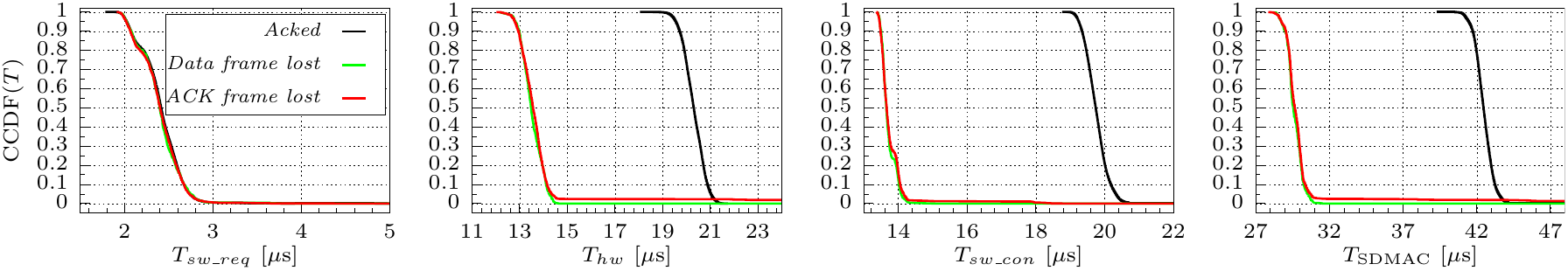}
  \else
  \includegraphics[width=2\columnwidth]{FIG6_TII-18-0703.pdf}
  \fi

  \caption{Complementary cumulative distribution functions of the different contributions to the SDMAC delay for sets $\mathcal{A}$, $\mathcal{L}_D$, and $\mathcal{L}_A$.}
  \label{plot:ccdfs}
\end{figure*}

\subsubsection{ACK timeout}
\label{sec:ACKTimeout}
Exact duration of frame transmission on air depends on its outcome.
In theory, the difference between durations of correctly performed and failed transmissions can be evaluated from equations \eqref{eq:TTr} and \eqref{eq:TTrs}, using values in Table~\ref{tab:parameters} and setting $T_{\operatorname{ACKTimeout}}$ to the default value $\unit[50]{\mu s}$ reported in the IEEE 802.11 specification \cite{2016-std-80211}, i.e., $\TTr-\TTr^* = \unit[10]{\mu s}$.
The same difference is expected when considering the time elapsing between timestamps $t^4$ and $t^5$, which corresponds to $\THw+\TTr$ and $\THwStar{}+\TTrStar{}$ values in Table~\ref{tab:outcomes},
for acked and non-acked frames, respectively.
By comparing set $\mathcal{A}$ to set $\mathcal{L}_D$, one can see that the measured difference is, on average, sensibly larger, i.e., $116.311-74.518=\unit[41.793]{\mu s}$.
Very similar differences can be obtained for $99$-percentiles, for which we have $117.287-75.424=\unit[41.863]{\mu s}$.
The main reason of this discrepancy is that, the actual $\emph{ACKTimeout}$ value set in the driver was lower than the default IEEE 802.11 value.
We verified that, for \mbox{Wi-Fi} adapters based on the \texttt{ath9k} driver, $T_{\operatorname{ACKTimeout}} = \unit[25]{\mu s}$. 
This is possibly due to the fact that such adapters have a short transmission range and are able to quickly detect the beginning of received frames.
For this reason, when computing the latency caused by the hardware on non-acked frames, $\THwStar = t^5 - t^4 - \TTrStar$, we set $\TTrStar=\unit[61]{\mu s}$ (while $\TTr=\unit[96]{\mu s}$ for acked ones).
From results in the table, the hardware appears to be slightly faster when no ACKs were received, i.e., $\THw - \THwStar \simeq \unit[7]{\mu s}$.

Differences were slightly larger for the overall path, where, on average, $T_{\mathit{path}}-T^*_{\mathit{path}}=138.483-90.628=\unit[47.855]{\mu s}$.
This is because the device-dependent code in the driver, which manages confirmations, is also slightly faster in case \emph{ACKTimeout} expired.
In particular, $\TSwConDRVdep - \TSwConDRVdepStar \simeq \unit[6]{\mu s}$.
Above results imply that, upon detection of a transmission failure, the application in the originator was notified, on average, $\unit[\sim\!\!48]{\mu s}$ in advance compared to successful transmissions.
This behavior can be profitably exploited by real-time applications running above SDMAC.
In fact, whenever a packet is lost, more time is given to the software to react to the event.
Such additional time can be used, e.g., to execute complex scheduling algorithms or heuristics aimed at counteracting the problem (e.g., by selecting another frame to be transmitted, changing the transmission rate, and so on).

Latencies for frames belonging to set $\mathcal{L}_A$ are similar, on average, to those of set $\mathcal{L}_D$.
Differences are described below.

\subsubsection{DATA frame loss vs. ACK frame loss}
Results for non-acked frames are reported in the two rightmost parts in Table~\ref{tab:outcomes}, for sets $\mathcal{L}_D$ and $\mathcal{L}_A$, respectively. 
Statistics are, on average, similar. 
In fact, from the point of view of the originator, the loss of either a DATA frame or the related ACK are almost indistinguishable, since both conditions are detected following the lack of the ACK frame.
The most significant differences regard standard deviation, higher-order percentiles, and maximum. 
In the case of set $\mathcal{L}_D$, when the DATA frame is lost (because of either corruption or collision), the related ACK frame is not returned by the recipient and, upon \emph{ACKTimeout} expiry, the originator detects that the transmission has failed.

Most of the transmissions in set $\mathcal{L}_A$ experienced an \emph{ACKTimeout} event, as for $\mathcal{L}_D$. 
According to the IEEE 802.11 specification, this occurs when ACK frame reception does not start within $T_{\operatorname{ACKTimeout}}$, e.g., because the PHY preamble is corrupted.
However, in a few cases the ACK frame did indeed arrive to the originator, but it was corrupted. 
In such event, the interrupt is raised by the \mbox{Wi-Fi} adapter after the ACK frame has been completely read and the relevant frame check sequence (FCS) verified.
Hence, failure is notified later, more or less at the same time as it would be in case of transmission success (see, e.g., the worst-case latencies $T_{\operatorname{max}}$ for $\mathcal{A}$ and $\mathcal{L}_A$, which are similar).
From the point of view of applications, these events (which actually depend on what takes place on air) can be suitably modeled as jitters introduced by SDMAC.
As a consequence, the distribution of latencies for $\mathcal{L}_A$ becomes bimodal, and higher-order percentiles and maximum are higher than $\mathcal{L}_D$.
For the same reason, standard deviation of $\mathcal{L}_A$ is noticeably larger than both $\mathcal{A}$ and $\mathcal{L}_D$.

\subsubsection{Validation of results}
Starting from the sample standard deviation and the number of samples included in each set, 
confidence intervals for mean values can be easily computed. 
Concerning the mean latency $\overline T_{\operatorname{SDMAC}}$, the confidence interval for set $\mathcal{A}$, which includes more than $12$ million samples, is $\unit[42.483 \pm 0.0004]{\mu s}$.
Reliability for $\mathcal{L}$ sets is not as good.
In particular, for $\mathcal{L}_D$, which includes $529$ samples, the confidence interval is $\unit[29.628 \pm 0.045]{\mu s}$, while for $\mathcal{L}_A$, which includes $1938$ samples but is characterized by a noticeably larger variance, it is $\unit[30.353 \pm 0.228]{\mu s}$.
Such intervals are narrow enough for the purposes of our analysis.

In addition, the complementary cumulative distribution functions (CCDFs) of the most important contributions to the measured latency $\TPc$ are reported in Fig.~\ref{plot:ccdfs} for the three sets of samples.
They provide a further confirmation of actual SDMAC determinism, and clearly highlight the differences between latencies of set $\mathcal{A}$ and those of sets $\mathcal{L}_D$ and $\mathcal{L}_A$.
In particular, the leftmost plot confirms that latencies in the request path do not depend in any way on the transmission outcome.
Similarly, the slight difference between $\mathcal{L}_D$ and $\mathcal{L}_A$ in the rightmost plot 
corroborates the hypothesis that the probability density function for the latter is bimodal.

\subsection{Applicability of SDMAC}
\label{sec:applicability}

\begin{figure}[b]
\centering
\includegraphics[width=1\columnwidth]{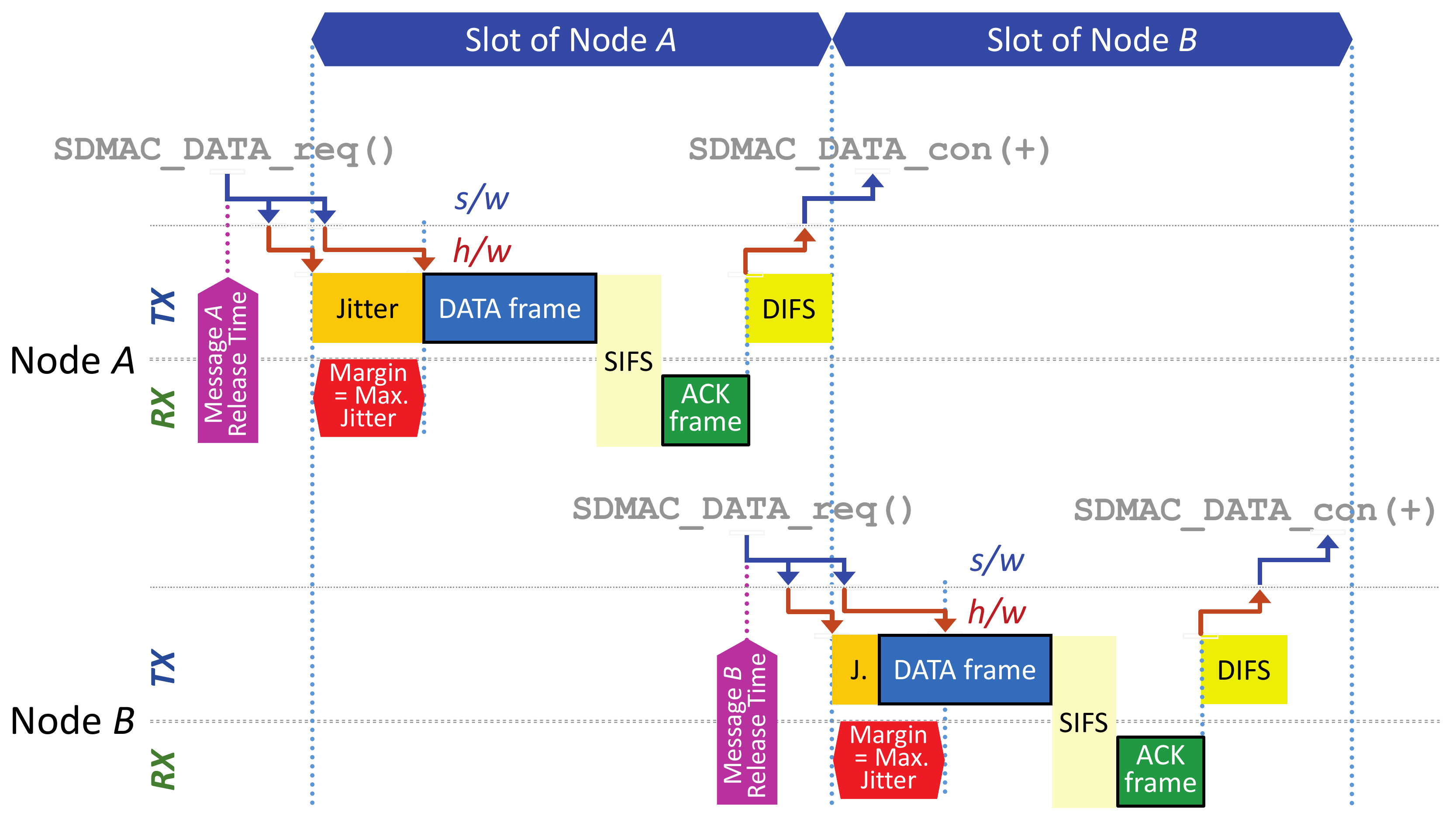}
\caption{Example of TDMA management using SDMAC.}
\label{fig:sdmacTDMA}
\end{figure}

Time-sensitive application scenarios listed in Section \ref{sub:application_contexts} can be analyzed in the light of the above results.

\subsubsection{Time Division Multiple Access}
Safety margins width is directly related to SDMAC determinism and synchronization quality between clocks of nodes.
As shown in Fig.~\ref{fig:sdmacTDMA}, if only jitters due to SDMAC were taken into account, the frame release jitter on air would be $J_{\operatorname{max}} = \operatorname{max}(\TSDMACReq)-\operatorname{min}(\TSDMACReq)$.
Since $\TSDMACReq$ is not made available by our testbed, its upper bound $\hat{T} = \TSwReq + \THw$ can be used, which realistically implies $J_{\operatorname{max}} \leq \hat{T}_{\operatorname{max}}-\hat{T}_{\operatorname{min}}$.

In the following we refer to percentiles, as they are more statistically reliable than worst-case values.
From measured delays in Table~\ref{tab:outcomes} for acked frames, a margin equal to $\hat{T}_{p99.9}-\hat{T}_{\operatorname{min}}=\left(3.797+21.455\right)-\left(1.786+18.051\right)= \unit[5.415]{\mu s}$ is large enough to guarantee that, reasonably, only one frame out of $1000$ falls out of time slot boundaries. 
Concerning clock synchronization over \mbox{Wi-Fi}, the $99.9$-percentile of the measured synchronization error for an implementation based on commercial PCs \cite{2015-TII-RBIS} is $\epsilon_{p99.9}=\unit[7.110]{\mu s}$, also including scheduling jitters (which represent the main contribution to uncertainties). 
As a very rough approximation, setting the safety margin to $5.415+7.110=\unit[12.525]{\mu s}$ should be enough to ensure a similar probability that slot boundaries are not exceeded at runtime.

\subsubsection{Deadline-driven traffic scheduling} 
SDMAC overhead is, on average, $\unit[42.483]{\mu s}$ for successful transmission attempts and $\unit[\sim\!30]{\mu s}$ for failed ones, which are comparable to the typical CSMA/CA timings (i.e., $T_{\mathit{DIFS}}=\unit[34]{\mu s}$ and $T_{slot}=\unit[9]{\mu s}$ when operating in the $\unit[5]{GHz}$ band).
Faster notification of failures (compared to the latencies experienced for successful attempts) and the ability to provide user space applications with statistics about the quality of the communication channel represent further advantages that justify SDMAC adoption in these application contexts.

\subsubsection{Seamless redundancy}
Above considerations still apply, as reasoning can be easily extended to cases where the same packet is concurrently sent on multiple adapters.

\subsubsection{Additional application contexts}
SDMAC applicability to other scenarios is part of our future work.
Its extensive performance characterization on a real prototype is, probably, the most important milestone of this paper, and proves that a variety of time-sensitive wireless systems can be implemented with this paradigm.

\section{Conclusions}
\label{sec:conclusion}
The SDMAC paradigm consists of a software overlay aimed at providing researchers, developers, and final users with full control on \mbox{Wi-Fi} adapters from user space applications. 
By relying on few and well-defined modifications to drivers, it enables effortless integration in existing commercial equipment, simple updating to newer driver releases, and supports easy customization to the features required by applications.

A number of guidelines are introduced and discussed in the paper, aimed at optimizing SDMAC.
Results obtained this way are quite promising:
for example, an experimental campaign which lasted $7$ days shows that latencies in our testbed, due to the software and hardware components of SDMAC, are bounded to reasonably low values, with statistically low jitters.
Despite the resulting system can not be considered hard real-time, all measured values for the SDMAC delay ranged from $\unit[39]{\mu s}$ to $\unit[77.5]{\mu s}$, and the 99.99-percentile is about $\unit[54]{\mu s}$.

The thorough performance evaluation we carried out evidences that SDMAC behavior is deterministic enough to make it an attractive enabling software component for many soft real-time applications, not only in industrial contexts.

\bibliographystyle{IEEEtran}
\bibliography{TII-18-0703}

\begin{IEEEbiography}%
[{\includegraphics[width=1in,height=1.25in,clip,keepaspectratio]
{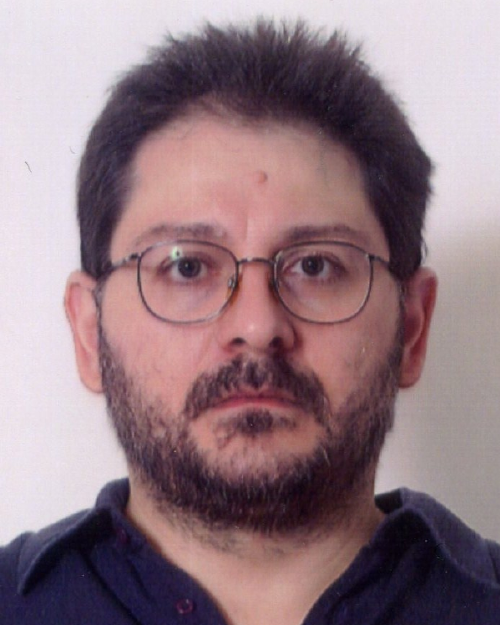}}]{Gianluca Cena} (SM'09) received the Laurea degree in electronic engineering and the Ph.D. degree in information and system engineering from the Politecnico di Torino, Italy, in 1991 and 1996, respectively. Since 2005 he has been a Director of Research with the Institute of Electronics, Computer and Telecommunication Engineering, National Research Council of Italy (CNR--IEIIT), Torino.

His research interests include wired and wireless industrial communication systems, real-time protocols, and automotive networks. In these areas he has coauthored about 130 technical papers, three of which awarded as Best Papers of the 2004, 2010, and 2017 editions of the IEEE Workshop on Factory Communication Systems, and one as 2017 Best Paper for the \textsc{IEEE Transactions on Industrial Informatics}, plus one international patent.

Dr. Cena served as a Program Co-Chairman for the 2006 and 2008 editions of the IEEE International Workshop on Factory Communication Systems, and as a Track Co-Chairman in six editions of the IEEE International Conference on Emerging Technologies and Factory Automation. Since 2009 he has been an Associate Editor of the \textsc{IEEE Transactions on Industrial Informatics}.
\end{IEEEbiography}

\begin{IEEEbiography}%
[{\includegraphics[width=1in,height=1.25in,clip,keepaspectratio]
{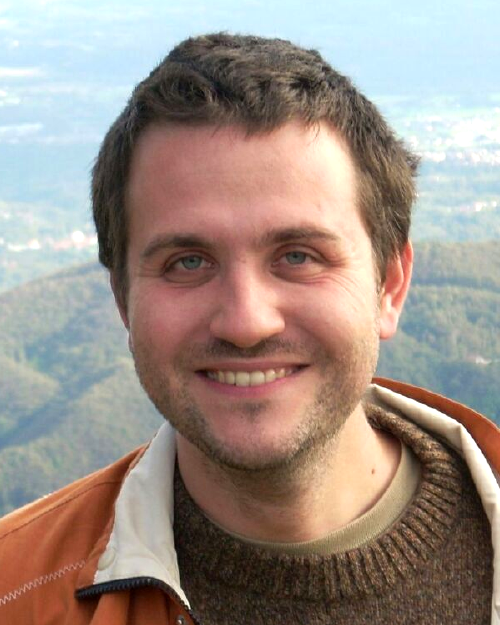}}]{Stefano Scanzio} (S'06-M'12) received the Laurea and Ph.D. degrees in Computer Science from Politecnico di Torino, Torino, Italy, in 2004 and 2008, respectively.
He was with the Department of Computer Engineering, Politecnico di Torino, from 2004 to 2009, where he was involved in research on speech recognition and, in particular, he has been active in classification methods and algorithms. Since 2009, he has been with the National Research Council of Italy (CNR), where he is a tenured Researcher with the Institute of Electronics, Computer and Telecommunication Engineering (IEIIT), Torino.

Dr. Scanzio served as a Work-in-Progress Co-Chairs in the 2018 edition of the IEEE International Workshop on Factory Communication Systems (WFCS 2018). He teaches several courses on Computer Science at Politecnico di Torino. He has authored and co-authored of more than 50 papers in international journals and conferences, in the area of industrial communication systems, real-time networks, wireless networks and clock synchronization protocols. He received the award for the best paper published in the \textsc{IEEE Transactions on Industrial Informatics} during 2016, and the Best Paper Awards for the papers he presented at the 8th and 13th IEEE Workshops on Factory Communication Systems (WFCS 2010 and WFCS 2017).
\end{IEEEbiography}

\begin{IEEEbiography}%
[{\includegraphics[width=1in,height=1.25in,clip,keepaspectratio]
{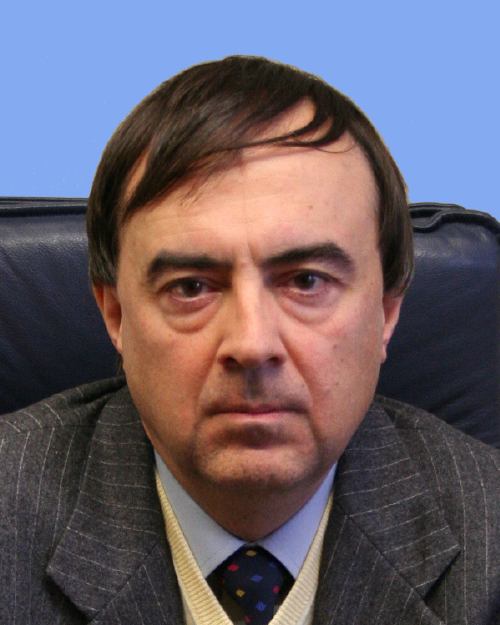}}]{Adriano Valenzano} (SM'09) received the Laurea degree magna cum laude in electronic engineering from Politecnico di Torino, Torino, Italy, in 1980. He is Director of Research with the National Research Council of Italy (CNR). He is currently with the Institute of Electronics, Computer and Telecommunication Engineering (IEIIT), Torino, Italy, where he is responsible for research concerning distributed computer systems, local area networks, and communication protocols. He has coauthored approximately 200 refereed journal and conference papers in the area of computer engineering.

Dr. Valenzano is the recipient of the 2013 IEEE IES and ABB Lifetime Contribution to Factory Automation Award. He was also awarded for the best paper published in the \textsc{IEEE Transactions on Industrial Informatics} during 2016, and received the Best Paper Awards for the papers presented at the 5th, 8th and 13th IEEE Workshops on Factory Communication Systems (WFCS 2004, WFCS 2010 and WFCS 2017).

Adriano Valenzano has served as a technical referee for several international journals and conferences, also taking part in the program committees of international events of primary importance. Since 2007, he has been serving as an Associate Editor for the \textsc{IEEE Transactions on Industrial Informatics}.
\end{IEEEbiography}

\end{document}